\documentclass[prd,preprint,superscriptaddress,showpacs,byrevtex]{revtex4}
\usepackage{bm}
\def\fsl#1{\setbox0=\hbox{$#1$}           
   \dimen0=\wd0                                 
   \setbox1=\hbox{/} \dimen1=\wd1               
   \ifdim\dimen0>\dimen1                        
      \rlap{\hbox to \dimen0{\hfil/\hfil}}      
      #1                                        
   \else                                        
      \rlap{\hbox to \dimen1{\hfil$#1$\hfil}}   
      /                                         
   \fi}                                         %
\newcommand{\be}{\begin{equation}}
\newcommand{\ee}{\end{equation}}
\newcommand{\bea}{\begin{eqnarray}}
\newcommand{\eea}{\end{eqnarray}}
\newcommand{\beq}{\begin{equation}}
\newcommand{\eeq}{\end{equation}}
\newcommand{\beqs}{\begin{eqnarray}}
\newcommand{\eeqs}{\end{eqnarray}}

\begin{document}
\title{ Nucleon Mass Sum Rule Violation in QCD and Confinement }
\author{Gouranga C Nayak }\thanks{G. C. Nayak was affiliated with C. N. Yang Institute for Theoretical Physics in 2004-2007.}
\affiliation{ C. N. Yang Institute for Theoretical Physics, Stony Brook University, Stony Brook NY, 11794-3840 USA}
\date{\today}
\begin{abstract}
In this paper we show that for a nucleon at rest if the quarks and/or antiquarks are in motion inside the nucleon producing chromo-magnetic field then the mass sum rule in QCD is violated when the confinement potential energy at large distance $r$ rises linearly with $r$ (or faster). Hence we find that the mass of the nucleon at rest is not equal to the mass-energy of all the quarks plus antiquarks plus gluons inside the nucleon if there exists chromo-magnetic field inside the nucleon and the confinement potential energy at large distance $r$ rises linearly with $r$ (or faster).
\end{abstract}
\maketitle
\pagestyle{plain}

\pagenumbering{arabic}

\section{ Introduction }

In the renormalized QCD \cite{hv} the asymptotic freedom \cite{asf} occurs at short distance where perturbative QCD (pQCD) is applicable. However, this does not help to answer the question how the nucleon is made up from quarks/antiquarks and gluons because the formation of nucleon from quarks/antiquarks and gluons is a long distance phenomena where the renormalized pQCD is not applicable. The long distance physics in the renormalzied QCD can be described by non-perturbative QCD but the analytical study of the non-perturbative QCD is not known yet.

In addition to this we have not experimentally observed isolated quarks/antiquarks as quarks/antiquarks are confined inside the hadron. Hence although the nucleon is a physical observable in QCD but the quarks/antiquarks and gluons are not physical observable. Because of this we do not know the value of the mass of the quark although we know the value of the mass of the nucleon. The mass of the real gluon is zero and the gluon carries energy.

According to nucleon mass sum rule in QCD the mass of the nucleon $N$ at rest is the sum of the mass-energy of all the quarks plus antiquarks plus gluons inside the nucleon. The nucleon mass sum rule in QCD is mathematically expressed as \cite{jaf,xji}
\bea
m_N = <\int d^3r {\hat T}^{00}(t,r)>
\label{mh}
\eea
where $m_N$ is the mass of the nucleon at rest, the ${\hat T}^{00}(x)$ is the $00$ component of the gauge invariant energy-momentum tensor operator ${\hat T}^{\nu \lambda}(x)$ of all the quarks plus antiquarks plus gluons inside the nucleon in QCD and the expectation value of an operator ${\hat {\cal O}}$ is defined by
\bea
<{\hat {\cal O}}>=\frac{<N|{\hat {\cal O}}|N>}{<N|N>}
\label{ex}
\eea
where $|N>$ is the (physical) state of the nucleon $N$ at rest. Note that the expectation $<{\cal O}>$ in eq. (\ref{ex}) is in the full QCD which means the $<{\cal O}>$ is a non-perturbative quantity in the renormalized QCD which can not be studied by using perturbative QCD (pQCD) since the nucleon is at rest. 

The eq. (\ref{mh}) can be written as
\bea
m_N=\sum_q <{\hat E}_q>+\sum_{\bar q} <{\hat E}_{\bar q}>+\sum_g<{\hat E}_g>
\label{fnfi}
\eea
where ${\hat E}_q$, ${\hat E}_{{\bar q}}$, ${\hat E}_g$ are the gauge invariant energy operators of the quark, antiquark, gluon [see eqs. (\ref{qemf}-\ref{qene})] and the $\sum_q$, $\sum_{\bar q}$, $\sum_g$ represent sum over all the quarks, antiquarks, gluons inside the nucleon. 

The nucleon mass sum rule in eq. (\ref{mh}) is based on the assumption that $<\int d^3r {\hat T}^{00}(t,r)>$ is a conserved quantity in QCD. It is claimed in \cite{jaf} that $\int d^3r { T}^{0\mu}(t,r)$ is a conserved charge. However, as we will show in this paper, this is not true in QCD. This is because although the gauge invariant energy-momentum tensor ${ T}^{\nu \lambda}(x)$ in the classical Yang-Mills theory satisfies the continuity equation
\bea
\partial_\nu T^{\nu \lambda}(x)=0
\label{emts}
\eea
but if the boundary surface term does not vanish in the classical Yang-Mills theory then $\int d^3r { T}^{00}(t,r)$ is not a conserved quantity. Note that we have used hat on the operator in quantum theory.

In this paper we find that for a nucleon at rest if the quarks and/or antiquarks inside the nucleon are in motion producing non-vanishing chromo-magnetic field then $<\int d^3r {\hat T}^{00}(t,r)>$ is not a conserved quantity in QCD if the confinement potential energy at large distance $r$ rises linearly with $r$ (or faster).

We find in this paper that for a nucleon at rest if the quarks and/or antiquarks are in motion inside the nucleon producing chromo-magnetic field and if the confinement potential energy at large distance $r$ rises linearly with $r$ (or faster) then
\bea
m_N=\sum_q <{\hat E}_q>+\sum_{\bar q} <{\hat E}_{\bar q}>+\sum_g<{\hat E}_g>+<{\hat m}_{\rm flux}>
\label{fni}
\eea
which does not agree with eq. (\ref{fnfi}) where $<{\hat m}_{\rm flux}>$ is the non-zero energy flux. The gauge invariant definition of the energy flux $m_{\rm flux}$ in the Yang-Mills theory is given by eq. (\ref{efx}).

Hence from eq. (\ref{fni}) we find that for a nucleon at rest if a non-vanishing chromo-magnetic field is produced by the motion of quarks and/or antiquarks inside the nucleon and if the confinement potential energy at large distance $r$ rises linearly with $r$ (or faster) then the mass of the nucleon at rest is not equal to the mass-energy of all the quarks plus antiquarks plus gluons inside the nucleon as the missing mass-energy is carried by the energy flux.

We will provide a proof of eq. (\ref{fni}) in this paper.

The paper is organized as follows. In section II we discuss the energy conservation and the vanishing energy flux in the Dirac-Maxwell theory. In section III we discuss  the QCD at infinite distance and the classical Yang-Mills theory. In section IV we prove eq. (\ref{fni}). Section V contains conclusions.

\section{Vanishing Energy Flux in Dirac-Maxwell Theory and the Conservation of Energy }\label{dmq}

From the gauge invariant Noether's theorem in Dirac-Maxwell theory we find \cite{nke}
\bea
&&\frac{dE_{EM}}{dt}+\frac{dE_e}{dt}=- \int d^3x ~{\vec \nabla} \cdot \left[ {\vec E}(x)\times {\vec B}(x)+ { \psi}^\dagger(x) [i{\vec \nabla} -e{\vec A}(x)] \psi(x)\right]
\label{ce}
\eea
where ${\vec E}$ is the electric field, $\psi$ is the Dirac field of the electron, ${\vec B}$ is the magnetic field, $A^\nu$ is the electromagnetic potential and
\bea
E_{\rm EM}=\frac{1}{2}\int d^3x ~[{\vec E}^2(x) + {\vec B}^2(x)]
\label{emf}
\eea
is the gauge invariant energy of the electromagnetic field and
\bea
E_e = \int d^3x~ \psi^\dagger(x) [i \partial_0 -eA_0(x)] \psi(x)
\label{ene}
\eea
is the gauge invariant energy of the electron.

Since the electromagnetic potential $A^\nu(t,r)$ falls off as $\frac{1}{r}$ in Dirac-Maxwell theory we find the vanishing energy flux
\bea
\int d^3x ~{\vec \nabla} \cdot \left[ {\vec E}(x)\times {\vec B}(x)+ { \psi}^\dagger(x) [i{\vec \nabla} -e{\vec A}(x)] \psi(x)\right]=0.
\label{vef}
\eea
From eqs. (\ref{vef}) in (\ref{ce}) we find
\bea
&&\frac{d [E_e+E_{EM}]}{dt}=0
\label{me}
\eea
which means the energy of the electron plus the energy of the electromagnetic field is conserved if the volume is infinite.

\section{ QCD at Infinite Distance and Classical Yang-Mills Theory }\label{hs}

Note that if the boundary surface is at finite distance then the boundary surface term is non-zero in classical Maxwell theory and in QED. Similarly if the boundary surface is at finite distance then the boundary surface term is non-zero in classical Yang-Mills theory and in QCD. In section \ref{dmq} we saw that if the boundary surface is at infinite distance then the boundary surface term is zero in Dirac-Maxwell theory. Hence we need to find out whether the boundary surface term is zero or non-zero in QCD when the boundary surface is at infinite distance.

As far as the boundary surface term at infinity is concerned the QED predicts that the QED potential is the Coulomb potential which agrees with the Coulomb form of the potential predicted by the classical Maxwell theory \cite{nkm}. Hence a vanishing boundary surface term at infinity in classical Maxwell theory means a corresponding vanishing boundary surface term at infinity in QED. As shown in section III of \cite{nkm}, since the Yang-Mills theory was discovered by making analogy with the Maxwell theory by extending U(1) gauge group to SU(3) gauge group \cite{yme,kp,ke}, one finds that as far as the boundary surface term at infinity is concerned the QCD predicts the same form of the potential that is predicted by the classical Yang-Mills theory. Hence, since the Yang-Mills theory was discovered by making analogy with the Maxwell theory by extending U(1) gauge group to SU(3) gauge group \cite{yme,kp,ke}, one finds that a non-vanishing boundary surface term at infinity in classical Yang-Mills theory means a corresponding non-vanishing boundary surface term at infinity in QCD, see section III of \cite{nkm} for details.

\section{ Nucleon Mass Sum Rule Violation in QCD and Confinement }

Consider a system in the classical Yang-Mills theory consisting of quarks plus antiquarks and the Yang-Mills potential (the color potential) $A_\mu^b(x)$ where $b=1,...,8$ is the color index. From the gauge invariant Noether's theorem in the classical Yang-Mills theory we find \cite{ky}
\bea
&&\frac{dE_{YM}}{dt}+\sum_q \frac{dE_q}{dt}+\sum_{\bar q} \frac{dE_{\bar q}}{dt}=- \int d^3x ~{\vec \nabla} \cdot [ {\vec E}^b(x)\times {\vec B}^b(x)+ \sum_{ q}{ \psi}^\dagger_j(x) [\delta^{jk}i{\vec \nabla} +gT^b_{jk}{\vec A}^b(x)] \psi_k(x) \nonumber \\
&&+(antiquarks)]
\label{qce}
\eea
where ${\vec E}^b$ is the chromo-electric field, $\psi_j$ is the Dirac field of the quark, ${\vec B}^b$ is the chromo-magnetic field, ${\bar q}$ represents antiquark, $\sum_q$ represents sum over all the quarks in the system, $(antiquarks)$ term is the corresponding boundary surface term for the antiquarks and
\bea
E_{\rm YM}=\frac{1}{2}\int d^3x ~[{\vec E}^b(x) \cdot {\vec E}^b(x) + {\vec B}^b(x) \cdot {\vec B}^b(x)]
\label{qemf}
\eea
is the gauge invariant energy of the Yang-Mills field (color field) and
\bea
E_q = \int d^3x~ \psi^\dagger_j(x) [\delta^{jk}i \partial_0 +gT^b_{jk}A_0^b(x)] \psi_k(x)
\label{qene}
\eea
is the gauge invariant energy of the quark.

Note that since we have not experimentally observed isolated quarks and/or antiquarks we know that quarks and/or antiquarks are confined inside the hadrons. For confinement to happen one finds that the chromo-electric field (chromo-magnetic field) can not fall faster than $\frac{1}{r^{\frac{3}{2}}}$ \cite{nkm}. If the potential energy at large distance $r$ rises linearly with $r$ (or faster) then we find that the chromo-electric field can not fall faster than
$\frac{1}{r}$ \cite{nkm}. The quark in motion produces chromo-magnetic field \cite{kp}. If the quarks and/or antiquarks are in motion inside the nucleon producing  chromo-magnetic field and if the chromo-electric field and the chromo-magnetic field do not fall faster than $\frac{1}{r}$ then we find the non-vanishing boundary surface term
\bea
\int d^3x ~{\vec \nabla} \cdot [ {\vec E}^b(x)\times {\vec B}^b(x)+ \sum_{ q}{ \psi}^\dagger_j(x) [\delta^{jk}i{\vec \nabla} +gT^b_{jk}{\vec A}^b(x)] \psi_k(x) +(antiquarks)] \neq 0.
\label{nbt}
\eea
Note that color potential also plays an important role in QGP study at high energy heavy-ion colliders \cite{nn1,nn2,nn3,nn4}.

From eqs. (\ref{nbt}) and (\ref{qce}) we find
\bea
&&\sum_q \frac{dE_q}{dt}+\sum_{\bar q} \frac{dE_{\bar q}}{dt}+\frac{dE_{YM}}{dt}+\frac{dm_{\rm flux}}{dt}=0
\label{qcf}
\eea
where the gauge invariant energy flux $m_{\rm flux}$ in the Yang-Mills theory is given by
\bea
m_{\rm flux}=\int d^4x ~{\vec \nabla} \cdot [ {\vec E}^b(x)\times {\vec B}^b(x)+ \sum_{ q}{ \psi}^\dagger_j(x) [\delta^{jk}i{\vec \nabla} +gT^b_{jk}{\vec A}^b(x)] \psi_k(x)+(antiquarks)]\nonumber \\
\label{efx}
\eea
where $\int dt$ integration is indefinite integration.

Extending eq. (\ref{qcf}) to QCD to study the mass $m_N$ of the nucleon $N$ at rest we find
\bea
\sum_q <\frac{d{\hat E}_q}{dt}>+\sum_{\bar q} <\frac{d{\hat E}_{\bar q}}{dt}>+\sum_g<\frac{d{\hat E}_g}{dt}>+<\frac{d{\hat m}_{\rm flux}}{dt}>=0
\label{qcg}
\eea
where $<...>$ is defined in eq. (\ref{ex}), the hat means corresponding operators by replacing $\psi, {\bar \psi},A \rightarrow {\hat \psi}, {\hat {\bar \psi}},{\hat Q}$ where $Q_\nu^b$ is the gluon field and the $\sum_{q,{\bar q},g}$ means sum over all the quarks, antiquarks, gluons inside the nucleon. Note that each term in the left hand side of eq. (\ref{qcg}) is renormalized in the renormalized QCD. Similarly the fields ${\hat \psi}, {\hat {\bar \psi}},{\hat Q}$ are renormalized fields in the renormalized QCD. We have used ${\hat Q}$ instead of ${\hat A}$ for the gluon field as we use the notation $A$ for the background field and $Q$ for the gluon field in the background field method of QCD to prove factorization and renormalization in QCD at high energy colliders at all orders in coupling constant \cite{nka}.

As mentioned in section \ref{hs} if the boundary surface is at finite distance then the boundary surface term is non-zero in classical Yang-Mills theory and in QCD.
We saw in section \ref{hs} that a non-vanishing boundary surface term at infinity in the classical Yang-Mills theory means a corresponding non-vanishing boundary surface term at infinity in QCD. Since the non-vanishing boundary surface term at infinity in the classical Yang-Mills theory gives non-vanishing energy flux $m_{\rm flux}$ in eq. (\ref{efx}) we find that the corresponding non-vanishing boundary surface term at infinity in QCD gives the corresponding non-vanishing energy flux
\bea
<{\hat m}_{\rm flux}> \neq 0.
\label{efxq}
\eea
From eqs. (\ref{efxq}) and (\ref{qcg}) we find that $\sum_q <{\hat E}_q>+\sum_{\bar q} <{\hat E}_{\bar q}>+\sum_g<{\hat E}_g>+<{\hat m}_{\rm flux}>$ is the conserved quantity which means the mass $m_N$ of the nucleon $N$ at rest is given by
\bea
m_N=\sum_q <{\hat E}_q>+\sum_{\bar q} <{\hat E}_{\bar q}>+\sum_g<{\hat E}_g>+<{\hat m}_{\rm flux}>
\label{fnf}
\eea
which reproduces eq. (\ref{fni}).

Hence we find that for a nucleon at rest if the quarks and/or antiquarks inside the nucleon are in motion producing non-vanishing chromo-magnetic field then the mass sum rule in QCD is violated if the confinement potential energy at large distance $r$ rises linearly with $r$ (or faster).

\section{Conclusions}
In this paper we have shown that for a nucleon at rest if the quarks and/or antiquarks are in motion inside the nucleon producing chromo-magnetic field then the mass sum rule in QCD is violated when the confinement potential energy at large distance $r$ rises linearly with $r$ (or faster). Hence we have found that the mass of the nucleon at rest is not equal to the mass-energy of all the quarks plus antiquarks plus gluons inside the nucleon if there exists chromo-magnetic field inside the nucleon and the confinement potential energy at large distance $r$ rises linearly with $r$ (or faster).


\begin{thebibliography}{99}

\bibitem{hv} G. 't Hooft and M.J.G. Veltman, Nucl.Phys. B44 (1972) 189.

\bibitem{asf} D. J. Gross and F. Wilczek, Phys. Rev. Lett. 30 (1973) 1343; D. Politzer, Phys. Rev. Lett. 30 (1973) 1346.

\bibitem{jaf} R. L. Jaffe and A. Manohar, Nucl. Phys. B337 (1990) 509.

\bibitem{xji} X-D. Ji, Phys. Rev. Lett. 74 (1995) 1071.

\bibitem{nke} G. C. Nayak, JHEP 1803 (2018) 101.

\bibitem{nkm} G. C. Nayak, arXiv:1804.02712v1 [hep-ph].

\bibitem{yme} C. N. Yang and R. Mills, Phys. Rev. 96 (1954) 191.

\bibitem{kp} G. C. Nayak, JHEP 1303 (2013) 001.

\bibitem{ke} G. C. Nayak, Eur. Phys. J. C73 (2013) 2442.

\bibitem{ky} G. C. Nayak, arXiv:1802.07825v1 [hep-ph].

\bibitem{nn1} F. Cooper and G. C. Nayak, Phys. Rev. D73 (2006) 065005; G. C. Nayak and P. van Nieuwenhuizen, Phys. Rev. D 71 (2005) 125001; M. C. Birse, C-W. Kao and G. C. Nayak, Phys. Lett. B570 (2003) 171; G. C. Nayak and R. S. Bhalerao, Phys. Rev. C 61 (2000) 054907; G. C. Nayak {\it et al.}, Nucl. Phys. A687 (2001) 457. 

\bibitem{nn2} F. Cooper, C-W. Kao and G. C. Nayak, Phys. Rev. D66 (2002) 114016; G. C. Nayak, Annals Phys. 325 (2010) 682; arXiv:1705.04878 [hep-ph]; Eur. Phys. J.C59 (2009) 891; Phys. Lett. B442 (1998) 427; JHEP 9802 (1998) 005; Eur. Phys. J. C64 (2009) 73; JHEP 0906 (2009) 071; Phys. Rev. D 72 (2005) 125010; C-W. Kao, G. C. Nayak and W. Greiner, Phys. Rev. D66 (2002) 034017.

\bibitem{nn3} F. Cooper, M. X. Liu and G. C. Nayak, Phys. Rev. Lett. 93 (2004) 171801;  G. C. Nayak, M. X. Liu and F. Cooper, Phys. Rev. D68 (2003) 034003; G. C. Nayak, Annals Phys. 324 (2009) 2579; Annals Phys. 325 (2010) 514; Eur. Phys. J.C59 (2009) 715; G. C. Nayak and V. Ravishankar, Phys. Rev. C 58 (1998) 356; Phys. Rev. D 55 (1997) 6877.

\bibitem{nn4} F. Cooper, E. Mottola and G. C. Nayak, Phys. Lett. B555 (2003) 181; A. Chamblin, F. Cooper and G. C. Nayak, Phys. Rev. D69 (2004) 065010; Phys. Lett. B672 (2009) 147; Phys. Rev. D70 (2004) 075018; D. Dietrich, G. C. Nayak and W. Greiner, Phys. Rev. D64 (2001) 074006.

\bibitem{nka} G. C. Nayak, JHEP 1709 (2017) 090; Phys. Part. Nucl. Lett. 13 (2016) 417; Eur. Phys. J. Plus 133 (2018) 52; Phys. Part. Nucl. Lett. 14 (2017) 18; Eur. Phys. J. C76 (2016) 448; arXiv:1705.07913 [hep-ph]; J. Theor. Appl. Phys. 11 (2017) 275.


\end{thebibliography}
\end{document}